\documentclass[aps,prl,twocolumn,superscriptaddress,amssymb]{revtex4-1}

\usepackage{amsmath}

\usepackage{lipsum}

\usepackage{graphicx}
\usepackage{floatrow}

\newcommand{\cB}{\mathcal{B}}
\newcommand{\cC}{\mathcal{C}}
\newcommand{\cD}{\mathcal{D}}
\newcommand{\cG}{\mathcal{G}}
\newcommand{\cN}{\mathcal{N}}
\newcommand{\cO}{\mathcal{O}}

\newcommand{\cH}{\mathcal{H}}

\newcommand{\nn}{\nonumber}

\begin{document}

\title{
Holographic Duals of Argyres-Douglas Theories
}

\author{Ibrahima Bah}
\affiliation{Department of Physics and Astronomy, Johns Hopkins University, 3400 North Charles Street, Baltimore, MD 21218, USA}
\author{Federico Bonetti}
\affiliation{Mathematical Institute, University of Oxford, Woodstock Road, Oxford, OX2 6GG, UK}
\author{Ruben Minasian}
\affiliation{Institut de Physique Th\'{e}orique, Universit\'{e} Paris Saclay, CNRS, CEA,  F-91191, Gif-sur-Yvette, France}
\author{Emily Nardoni} 
\affiliation{Mani L. Bhaumik Institute for Theoretical Physics, Department of Physics and Astronomy, University of California, Los Angeles,  CA 90095, USA}

\begin{abstract}

We propose the first explicit  holographic duals for
a class of
superconformal field theories  of Argyres-Douglas type, which
are inherently
strongly coupled and provide a window onto remarkable non-perturbative
phenomena (such as mutually non-local massless dyons and relevant operators of fractional dimension).
The theories under examination are realized by a stack of M5-branes wrapped on a sphere with one irregular puncture and one regular puncture. 
In the dual 11d supergravity solutions,
the irregular puncture is   realized as  an internal M5-brane source.


%
%
%
%
%

\end{abstract}


\maketitle

\section{Introduction  \label{intro}}

Strong-coupling phenomena in quantum field theory (QFT)
are of crucial importance, both conceptually and phenomenologically,
but their study poses considerable theoretical challenges.
In the endeavor of exploring the vast and largely uncharted landscape of strongly
coupled phases in QFT, valuable lessons can be learned from theories
with a higher degree of symmetry.
Superconformal field theories (SCFTs) of Argyres-Douglas (AD) type in four dimensions
constitute a prominent example. These theories are intrinsically strongly-coupled and 
describe interactions among  mutually non-local massless dyons~\cite{Argyres:1995jj}. 
Their  spectrum contains relevant 
Coulomb branch operators of fractional dimension.
Establishing the existence and surprising properties of these QFTs has been complicated by their lack of an $\cN=2$ weak-coupling Lagrangian description, and hence, exploring less conventional windows into their physics is especially valuable.

%

A vast class of SCFTs of AD type is expected to admit holographic duals,
but their identification 
has remained an open problem for years. In this letter,
we present a new class of fully explicit $AdS_5$ solutions in 11d supergravity
and we propose them as 
 holographic duals to SCFTs of AD type.
Our results   give the opportunity to analyze these QFTs
from a new angle, providing novel insights on their properties.
Furthermore, a subclass of SCFTs of AD type
can be realized as  $\cN = 2$ supersymmetric IR fixed points of
renormalization group (RG) flows preserving $\cN = 1$ supersymmetry  \cite{Agarwal:2017roi,Benvenuti:2017bpg}.
Our solutions   pave the way to the exciting possibility of studying the gravity dual of   supersymmetry enhancing RG flows, which could shed new light on holography 
 in general.

A crucial feature of our $AdS_5$    solutions is 
the presence of suitable singularities, which we interpret as 
 the low-energy approximation to well-defined brane sources
in   M-theory.
Localized sources in the internal space constitute an important ingredient in the holographic dictionary that allows for arbitrary flavor symmetries (see \emph{e.g.}~\cite{Brandhuber:1999np, Apruzzi:2013yva, Gaiotto:2014lca, Apruzzi:2015wna, DHoker:2016ujz, DHoker:2017mds, Bah:2017wxp, Bah:2018lyv}). 
This letter describes novel controlled examples 
allowing  a better understanding of these sources, pivotal for enlarging the scope of the AdS/CFT
correspondence.

\section{Supergravity Solutions  \label{spindle}}

Our    $AdS_5$ solutions in 11d supergravity preserve 4d $\cN = 2$ superconformal symmetry.
They were  obtained in 7d gauged supergravity
and uplifted on $S^4$, as will be reported in \cite{toappear}.
The 7d solutions are a warped product of $AdS_5$ and a 2d
space $\Sigma$, consisting of a circle fibered over   an interval.
$\Sigma$ is supported by a $U(1)$ gauge flux, does not have a constant curvature metric, and admits a non-constant
Killing spinor. Thus, as in  \cite{Bah:2013wda,Ferrero:2020laf,Ferrero:2020twa},
supersymmetry is   not realized in  the standard topological twist paradigm.

The metric of   the uplifted 11d solution is
\begin{align}
ds^2_{11} &= m^{-2} \, e^{2\lambda} \, (ds^2_{AdS_5} + ds^2_{M_6}) \ , 
\label{general11d}\\
ds^2_{M_6} & = \frac{dw^2}{2 \, w\,  h(w) \, (1-w^2)^{3/2}} + \frac{\cC^2 \, h(w) \, dz^2}{B}
\label{our_M6}
   \\
&  \!\!\!\!\!\!\!\!\!\!\!\!\!\!\!  + \frac{\sqrt{1-w^2}}{2 \, B} \, \bigg[  \frac{d\mu^2}{w\,(1-\mu^2)}  + \frac{(1-\mu^2)\, D\phi^2}{w\, \cH(w,\mu)}\, 
+ \frac{w \, \mu^2 \, ds^2_{S^2}}{\cH(w,\mu)}  \bigg] \ , \nn
\end{align}
where $m$ is a mass scale, $ds^2_{ AdS_5}$ is the
metric on the unit-radius $AdS_5$, and $ds^2_{ S^2}$ is the
metric on the unit-radius $S^2$.
The functions $h(w)$, $\cH(w,\mu)$ are defined as
\begin{align}
h  = B - 2 \, w \, \sqrt{1-w^2} \ , \quad \cH  = \mu^2 + w^2 \, (1-\mu^2) \ , 
\end{align}
where $0<B<1$ is a constant parameter. 
The coordinates $\mu$, $w$  have ranges $0\le \mu \le1$ and $0 \le w \le w_1$, with
$w_1^2 = \frac 12 \, (1 - \sqrt{1-B^2})$.
The angular coordinates $\phi$, $z$ have period $2\pi$,
and $\cC$ is a constant.
The 1-form $D\phi$ and the warp factor are given by
\begin{align}
\!\!\! D\phi  = d\phi + \cC \, (2\,w^2-1) \, dz  \ , \quad
e^{2\lambda}  = \frac{2 \,B \, w^{1/3} \, \cH^{1/3}}{\sqrt{1-w^2}}  \ .
\end{align}
The $G_4$ flux supporting the solution reads
\begin{align}
G_4 = - \frac{1}{m^3} \, {\rm vol}_{S^2} \wedge d\bigg[ \frac{\mu^3 \, D\phi }{\cH} \bigg] \ ,
\end{align}
where  ${\rm vol}_{S^2}$ is the volume form
of the $S^2$. 

\begin{figure}
\centering

\includegraphics[width=7cm]{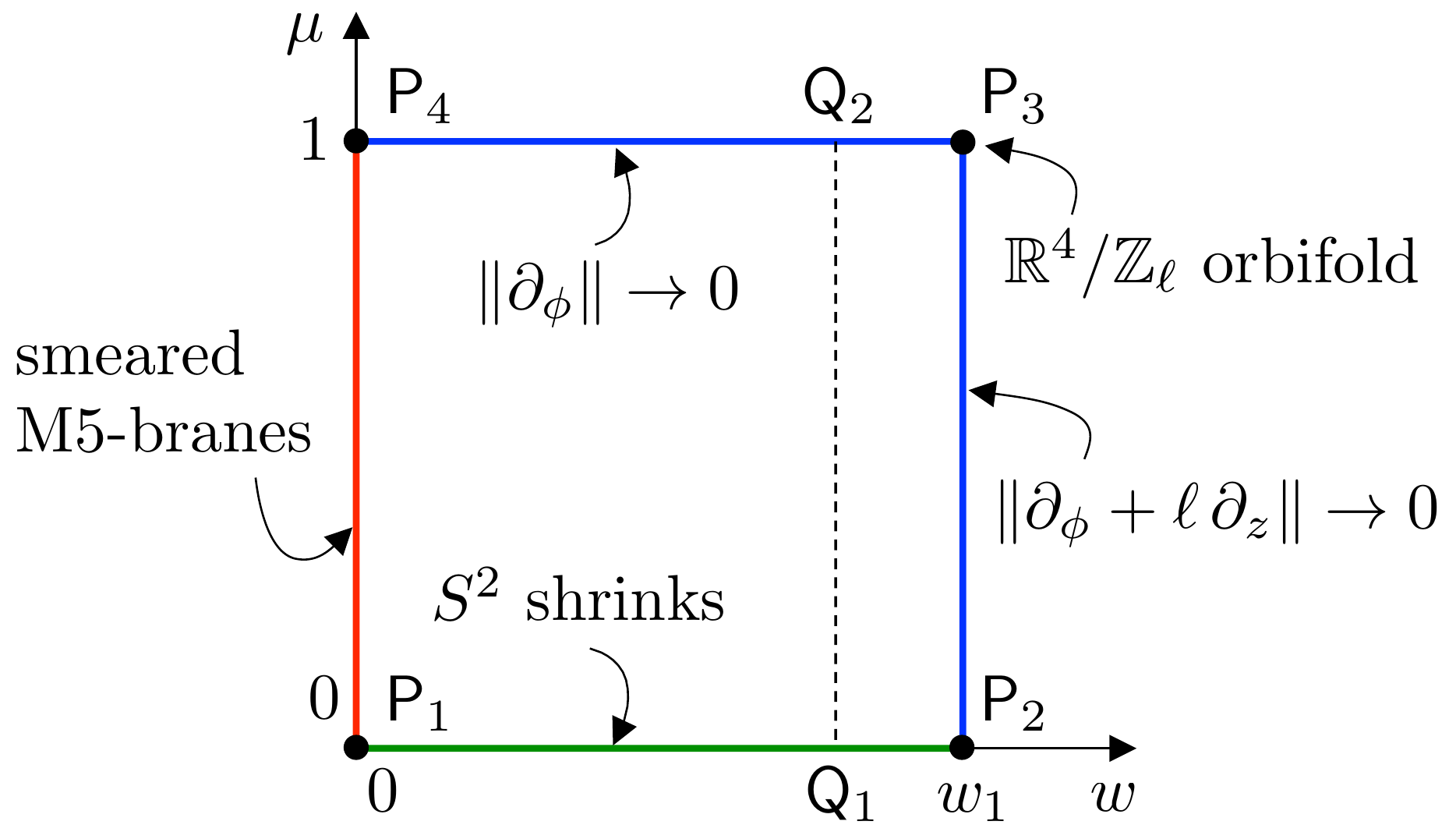}

     \caption{The internal space is an $S^1_\phi \times S^1_z \times S^2$ fibration
over   
$[0,w_1]\times [0,1]$
  in the $(w,\mu)$ plane. The $S^2$ shrinks smoothly along $\mathsf P_1\mathsf P_2$.
Different linear combinations
of $\partial_\phi$, $\partial_z$ shrink smoothly along $\mathsf P_2 \mathsf P_3$
and $\mathsf P_3 \mathsf P_4$, as indicated.
At $\mathsf P_3$ the 4d space parametrized by $w$, $\mu$, $\phi$, $z$ 
is locally $\mathbb R^4/\mathbb Z_\ell$.
The region near $\mathsf P_1 \mathsf P_4$    
is interpreted in terms of smeared M5-branes. The segment $\mathsf Q_1\mathsf Q_2$ enters
the definition of the 4-cycle~$\cC_4$.
\label{fig_geometry}}
\end{figure}

The space $M_6$ is an $S^1_z \times S^1_\phi \times S^2$
fibration over the rectangle $[0,w_1]\times [0,1]$ in the $(w,\mu)$ plane,
see Figure~\ref{fig_geometry}. 
The directions $w$, $S^1_z$ in \eqref{our_M6} are identified with $\Sigma$ in
the 7d  solution, while $\mu$, $S^1_\phi$, $S^2$
span the $S^4$ used in the uplift.

\subsection{Regularity and Flux Quantization}

As we approach a  point in the interior of the $\mathsf P_1 \mathsf P_2$
segment in the $(w,\mu)$ plane (see Figure \ref{fig_geometry}), the
$S^2$ shrinks smoothly.
The   Killing vector $\partial_\phi$ shrinks smoothly in the interior of 
$\mathsf P_3 \mathsf P_4$.
The linear combination $\partial_\phi + \ell \, \partial_z$ shrinks smoothly along $\mathsf P_2 \mathsf P_3$, where $\ell$ is given as
\begin{align} \label{ell_def}
\ell = \frac{1}{\cC \, \sqrt{1-B^2}} \ , \qquad \ell \in \mathbb N \ .
\end{align}
 The quantization of $\ell$ stems from
analyzing the local geometry of the 4d space spanned
by $w$, $\mu$, $\phi$, $z$ near $\mathsf P_3$, and requiring
it to be locally an orbifold $\mathbb R^4/ \mathbb Z_\ell$.

The internal space $M_6$ admits non-trivial
4-cycles which lead to flux quantization conditions for $G_4$.
The 4-cycle $\cC_4$ is obtained by combining the
segment $\mathsf Q_1 \mathsf Q_2$, $S^1_\phi$,
and $S^2$. $\cC_4$ has the topology of a 4-sphere
because the $S^2$ shrinks at $\mathsf Q_1$ and the $S^1_{\phi}$ shrinks at $\mathsf Q_2$.
The flux of $G_4$ through $\cC_4$ with suitable orientation
defines  
\begin{align} \label{N_def}
N = \int_{\cC_4}\frac{G_4}{(2\pi \ell_p)^3} = \frac{1}{\pi m^3 \ell_p^3} \ , \qquad N \in \mathbb N \ ,
\end{align}
where $\ell_p$ is the 11d Planck length.
Next, we define the 4-cycle $\cB_4$
by combining $S^2$, the segment $\mathsf P_2 \mathsf P_3$,
and the linear combination of $S^1_{\phi}$ and $S^1_z$
that does \emph{not} shrink in the interior of 
$\mathsf P_2 \mathsf P_3$. $\cB_4$ is topologically a 4-sphere,
because the $S^2$ shrinks at $\mathsf P_2$ and both $S^1_\phi$ and $S^1_z$
shrink at the orbifold point $\mathsf P_3$.
For $\ell = 1$ we have $\cB_4 \cong \cC_4$, but 
$\cB_4$ is an independent 4-cycle for $\ell >1$, with
\begin{align} \label{N_over_ell}
\mathbb Z \ni \int_{\cB_4}\frac{G_4}{(2\pi \ell_p)^3} = \frac{N}{\ell}   \ , \quad \text{hence 
 $\ell$ divides
 $N$.}
\end{align}
Finally, we construct the 4-cycle  $\cD_4$ 
by combining $\mathsf P_3 \mathsf P_4$ with $S^2$---which shrinks at $\mathsf P_4$---and the  combination of $S^1_{\phi}$ and $S^1_z$ that does \emph{not}
shrink in the interior of $\mathsf P_3 \mathsf P_4$.
Integrating $G_4$ on $\cD_4$ defines  
\begin{align} \label{K_def}
K =  \int_{\cD_4}\frac{G_4}{(2\pi \ell_p)^3}  =  \frac{N \, (1-\sqrt{1-B^2})}{\ell \,\sqrt{1-B^2}}  \ , \quad K \in \mathbb N \ .
\end{align}

In the vicinity of  $\mathsf P_1 \mathsf P_4$,
the geometry is singular and $e^{2\lambda}$ vanishes.
We interpret this in terms of a smeared M5-brane source, as inferred from 
 $G_4$ near $w = 0$,
\begin{align} \label{local_G4}
G_4 =   - \frac{{\rm vol}_{S^2} \wedge d \mu \wedge D\phi }{m^3} + \dots \ .
\end{align}
This term yields a finite flux equal to $N$  \eqref{N_def} when integrated along $S^2$, $\mu$, $S^1_\phi$, signaling a source of the form $dG_4 \sim \delta(w) \, dw \wedge {\rm vol}_{S^2} \wedge d \mu \wedge D\phi$.
Comparing the 11d metric near $w=0$ with the standard M5-brane solution,
we see that the
M5-branes  are extended along $AdS_5$ and the $S^1_z$,
smeared along the $\mu$, $S^1_\phi$ directions, and sitting at the origin $w=0$
of the local $\mathbb R^3$ space~$dw^2 + w^2 \, ds^2_{S^2}$.

\subsection{Solutions in Canonical $\mathcal{N}=2$ Form}

The general form of an  $AdS_5$ M-theory solution preserving
4d $\cN = 2$ superconformal symmetry
was determined in \cite{Lin:2004nb} by
Lin, Lunin, and Maldacena (LLM).
The   11d metric and flux are summarized in \cite{Gaiotto:2009gz}.
In LLM form, the internal space $M_6$ is
an $S^1_\chi \times S^2$ fibration over a 3d space
with local coordinates $(x_1,x_2,y)$.
The Killing vector $\partial_\chi$ is associated with the 
$U(1)_r$ R-symmetry of the dual SCFT,
while the isometries of the $S^2$ are mapped to the $SU(2)_R$
R-symmetry. The solution is determined by
a  function
$D(x_1, x_2, y)$ satisfying the Toda  equation
\begin{align}
(\partial_{x_1}^2  + \partial_{x_2}^2 ) D + \partial_y^2 e^D = 0 \ . \label{toda}
\end{align}
Our solutions can be cast in canonical LLM form, with
the $S^2$ in \eqref{our_M6} identified with the $S^2$ in LLM.
Defining $x_1 + i \, x_2 = r \, e^{i \beta}$, 
the map between  $\chi$, $\beta$
and  $\phi$, $z$ is
\begin{align} \label{chi_and_beta}
\!\! \!  \partial_\chi = \partial_\phi  \!+ \! \frac{N \, \ell }{N +  K  \ell }    \partial_z   \, ,   \;\;
\partial_\beta = \partial_\phi \!+ \!\left[  1 +  \frac{N \, \ell }{N +  K  \ell }  \right] \! \partial_z   \, . \! \!
\end{align}
With reference to the uplift from 7d, the isometry $\partial_\chi$ 
mixes the  $\Sigma$ and $S^4$ directions. This is
in contrast to the solutions of  \cite{Maldacena:2000mw, Gaiotto:2009gz},
in which
$\partial_\chi = \partial_\phi$.

The function $D$ and the map between the LLM coordinates $y$, $r$
and the coordinates $\mu$, $w$, are
\begin{align}
y &= \frac{4\, B \, w\, \mu}{  \sqrt{ 1-w^2  }  } \ , \qquad
r = (1-\mu^2)^{-\frac{1}{2\cC} } \, \cG(w) \ ,   \\
e^D &= \frac{16\, B  \,  \cC^2 \,  \left(1-\mu ^2\right)^{1 + \frac 1 \cC } \, h}{     \left(1 - w^2\right)   \, \cG(w)^2 }  \ ,
\;\;
\frac{\cG'(w)}{\cG(w)} = \frac{-B \, w  }{\cC   
\left(1-w^2 \right)  h}  \ .\nn 
\end{align}
This determines a class of exact solutions
$D$ to the Toda equation \eqref{toda} which are separable in the variables
$\mu$, $w$. Crucially, in our setup $D$ does not describe a constant
curvature Riemann surface, in contrast to the 4d $\cN = 2$ Maldacena-Nu\~nez solutions
\cite{Maldacena:2000mw}.

\subsection{Holographic Central Charge, Flavor Central Charge, and Probe M2-Branes \label{ccharge}}

The holographic central charge
is extracted from the warped volume of the internal space \cite{Gauntlett:2006ai},
\begin{align} \label{holo_c}
c = \frac{1}{2^7   \pi^6 m^9 \ell_p^9} \,\int_{M_6} e^{9   \lambda} \,  {\rm vol}_{M_6}
= \frac{\ell \, N^2 \, K^2}{12 \, (N +  K\, \ell)} \ ,
\end{align}
where ${\rm vol}_{M_6}$ is the volume form of    $ds^2_{M_6}$
in \eqref{general11d}. 

Expanding the M-theory 3-form $C_3$ onto the
resolution cycles of the $\mathbb R^4/\mathbb Z_\ell$ orbifold singularity at
$\mathsf P_3$, one obtains $\ell-1$ Abelian gauge fields.
The gauge group enhances to $SU(\ell)$ by virtue of
states from M2-branes wrapping the resolution cycles \cite{Gaiotto:2009gz}.
We compute the associated flavor central charge $k_{SU(\ell)}$ 
using the 't Hooft anomaly inflow methods of \cite{Bah:2019rgq}, yielding
\begin{align} \label{holo_flavor_charge}
k_{SU(\ell)}   = \frac{2 \, N \, K \, \ell}{N + K \, \ell} \ .
\end{align}

M2-brane probes wrapping calibrated 2-cycles in the internal space
are dual to BPS operators in the SCFT. 
The calibration condition 
was written in \cite{Gauntlett:2006ai} for a generic solution preserving 
4d $\cN = 1$ superconformal symmetry and can be 
adapted to the $\cN = 2$ solutions at hand.
The conformal dimension $\Delta$ of the operator
dual to an M2-brane wrapping the calibrated 2-cycle $\cC_2$
is \cite{Gauntlett:2006ai}
\begin{align} \label{dimension_formula}
\Delta = \frac{1}{4  \pi^2  m^3 \ell_p^3} \, \int_{\cC_2} \, e^{3   \lambda} \,  {\rm vol}_{\cC_2}   \ ,
\end{align}
where $ {\rm vol}_{\cC_2}$ is the volume form on $\cC_2$ induced
from $ds^2_{M_6}$.

We identify two supersymmetric M2-brane probes in our setup.
Firstly, we can wrap an M2-brane on 
the $S^2$ on top
of the orbifold point $\mathsf P_3$ in the $(w,\mu)$ plane.
We denote the corresponding operator as $\cO_1$.
Secondly, we can wrap an M2-brane on the 2d subspace 
consisting of
 the segment $\mathsf P_3 \mathsf P_4$
and the combination of $S^1_\phi$ and $S^1_z$ that does \emph{not} shrink
in the interior of $\mathsf P_3 \mathsf P_4$. 
This subspace
corresponds to an open M2-brane 
ending on the M5-branes at $w=0$.
We denote the associated operator as $\mathcal O_2$.
The dimensions of $\cO_1$, $\cO_2$~from~\eqref{dimension_formula}~are 
\begin{align} \label{wrapped_dim}
\Delta(\cO_1)  = \frac{N \, K \, \ell}{N + K \, \ell} \ , \qquad \Delta(\cO_2) = K \ .
\end{align}
The $U(1)_r \times SU(2)_R$ charges of $\cO_1$, $\cO_2$
can be computed from the M2-brane coupling to the 11d 3-form $C_3$  \cite{Gauntlett:2006ai},
\begin{align} \label{wrapped_charges}
\! \!   (r,R)(\cO_1)  \! = \!   (2\Delta(\cO_1) ,0 ) \, , \,\,
(r,R)(\cO_2)  \! = \!  ( 0, \Delta(\cO_2) ) \, ,
\end{align}
with $R$
  the Cartan generator  of 
$SU(2)_R$. Thus $\cO_1$ and $\cO_2$ have the R-charges of $\mathcal{N}=2$ Coulomb branch and Higgs branch operators, respectively.

\section{A Novel St{\"u}ckelberg Mechanism  \label{ccharge}}

The Killing vector $\partial_\beta$ in \eqref{chi_and_beta} 
is a symmetry of the 11d metric and flux,
but it does not correspond to a continuous flavor symmetry
of the dual SCFT. This is due to a  
St{\"u}ckelberg mechanism in the 5d low-energy
effective action of M-theory on $M_6$.
The components of the 11d metric with one external
leg and one leg along $\partial_\beta$ yield
a $U(1)$ gauge field $A^\beta$. 
When $A^\beta$ is turned on,
the 1-form $d\beta$ must be replaced by the 
gauge invariant combination $d\beta + A^\beta$.
This replacement affects the closure of $G_4$,
which is restored by adding suitable terms, including
\begin{align}
G_4^{\rm tot} = G_4|_{d\beta \rightarrow d\beta +A^\beta} + Da_0 \wedge \omega_3 + \dots \ .
\end{align}
The improved $G_4^{\rm tot}$ is built with 
the closed but not exact 3-form 
 $\omega_3 \propto \iota_{\partial_\beta} G_4$, whose
  non-exactness
hinges on the   M5-brane source at $w=0$. 
The 1-form $Da_0$ is the field strength
of an external axion $a_0$. 

Closure of $G_4$ requires
$dDa_0 \propto dA^\beta$, signaling a non-trivial
St{\"u}ckelberg coupling between $A^\beta$ and $a_0$.
As a result, $A^\beta$ is massive  
and is dual to a spontaneously broken $U(1)$ symmetry in the SCFT.
As discussed in detail in \cite{Bah:2021hei},
this mechanism provides a non-trivial physical realization 
of a mathematical  
obstruction to promoting  $G_4$  to an equivariant
cohomology class \cite{WU1993381}.  
In contrast, 
$ \iota_{\partial_\chi} G_4$ is exact, and
the  $U(1)$ gauge field $A^\chi$ 
(originating from the components of the 11d metric with one external leg and one  leg along $\partial_\chi$)
does not participate in 
any St{\"u}ckelberg coupling to $a_0$ and remains massless. This is expected
since $\partial_\chi$ is dual to the $U(1)_r$ R-symmetry of the SCFT.
Similar
versions of the St\"uckelberg mechanism
for isometries in flux backgrounds are known for flat internal spaces
(see \emph{e.g.}~\cite{Hinterbichler:2013kwa}). The internal geometry
discussed in this letter is richer, and this St\"uckelberg mechanism
is novel in the context of holographic M-theory solutions.

\section{Field Theory Duals  \label{ft}}

We claim that the 11d supergravity solutions presented above are holographically
dual to 4d $\mathcal{N}=2$ SCFTs 
 that arise from the low energy limit of $N$ M5-branes wrapping a sphere with an irregular puncture of type $A_{N-1}^{(N)}[k]$, labeled by the integer $k > -N$. For $\ell =1$ the irregular puncture is the only puncture on the sphere, and the 4d field theories coincide with the Type I theories with $b=N$ and $J=A_{N-1}$ in the classification of \cite{Xie:2012hs,Wang:2015mra} (also called $I_{N,k}$ in \cite{Xie:2013jc}). These are the  AD theories of type $(A_{N-1},A_{k-1})$,
obtained in Type IIB in \cite{Cecotti:2010fi} (generalizing the $N=2$ cases  obtained  in \cite{Argyres:1995jj,Argyres:1995xn,Eguchi:1996vu}).  
  For $\ell > 1$ there is an additional regular puncture at the opposite pole of the sphere that is labeled by a box Young diagram with $\ell$ columns and $N/\ell$ rows, contributing an $SU(\ell)$ non-Abelian flavor symmetry \footnote{Box Young diagrams from orbifold singularities were studied in \cite{Bobev:2019ore}.}. We label the resulting 4d theories by $(A_{N-1}^{(N)}[k],Y_{\ell})$, which belong to the class labeled Type IV in  \cite{Xie:2012hs,Wang:2015mra}. 
For $\ell = N$ the regular puncture is of maximal type and these
are the $D_{k+N}^{N}(SU(N))$ theories studied in \cite{Cecotti:2012jx,Cecotti:2013lda,Giacomelli:2017ckh}. The case $\ell=1$ is the ``non-puncture", equivalent to the $(A_{N-1},A_{k-1})$ theories.

The irregular puncture is identified with the M5-brane source in the gravity dual. Due to the  irregular puncture, the $U(1)_r$ R-symmetry of the SCFT is given as the combination $r = R_\phi + \frac{N}{k+N} R_z$, where $R_\phi$ is the generator of the R-symmetry that  would be preserved in the absence of the irregular defect and $R_z$ is the generator of the global $U(1)_z$ isometry of the sphere \cite{Xie:2012hs,Wang:2015mra}. 
Comparison with \eqref{chi_and_beta} gives the map between  $k$ in the SCFT and the flux quantum $K$,
\begin{align} \label{K_vs_k}
K = k  + N \left( 1- \tfrac{1}{\ell} \right)\ .
\end{align}

The central charges of the $(A_{N-1}^{(N)}[k],Y_{\ell})$ theories are summarized in Table \ref{tab:tp1}.
They are computed in the literature \cite{Xie:2012hs,Cecotti:2013lda,Xie:2015rpa,Xie:2016evu,Giacomelli:2017ckh} using useful formulae from \cite{Shapere:2008zf}. For $\ell > 1$, an especially simple way to compute the central charges as a function of $\ell$ is to apply the results of \cite{Giacomelli:2017ckh, Giacomelli:2020ryy} for the partial closure of a maximal puncture, initiated by a nilpotent VEV for the moment map operator of the maximal puncture's flavor symmetry. 
The third row of Table \ref{tab:tp1} 
gives the central charge in the limit $N,k \rightarrow \infty$  with
$k/N$ finite.
Using \eqref{K_vs_k}, we  get a perfect 
match with 
the holographic central charge \eqref{holo_c}.

The dimensions of the  Coulomb  branch operators $u_i$
of the theory  $(A_{N-1}^{(N)}[k],Y_{\ell})$ 
are conveniently captured by a Newton polygon \cite{Xie:2012hs}
and obey the bounds
	\begin{align}\label{eq:delt} 
	1 < \Delta(u_i) \leq  N   - \frac{ N^2}{\ell \, (N+k)}  \ .
	\end{align}
The upper bound is saturated by exactly one $u_i$,
which     has the correct dimension and R-charges to
be identified with the    M2-brane operator $\cO_1$
in~\eqref{wrapped_dim},~\eqref{wrapped_charges}~\footnote{A Coulomb branch operator has $U(1)_r$ charge $r = 2\Delta$ and is uncharged under $SU(2)_R$.}.

Using \eqref{K_vs_k}, the $k_{SU(\ell)}$ central charge \eqref{holo_flavor_charge} reads
\begin{align}
k_{SU(\ell)} = 2 \,N   - \frac{2 \, N^2}{\ell \, (N+k)} \ .
\end{align}
For $\ell = N$ it matches the field theory computation of~\cite{Cecotti:2013lda}.
For generic $\ell$, it matches the conjecture of \cite{Xie:2013jc}
that the flavor central charge is equal to twice the maximal Coulomb branch operator dimension---see \eqref{eq:delt}.

For $\ell =1$, the rank of the global symmetry of the $(A_{N-1},A_{k-1})$ theories is  $\text{GCD}(k,N)-1$ \cite{Giacomelli:2017ckh}.   
The maximal rank $N-1$ on the SCFT side matches with the
maximal rank that can be achieved via the M5-brane source on the gravity side.
It would be interesting to establish a  precise match with the SCFT formula for generic $k$, $N$.
 
\renewcommand{\arraystretch}{2}
\begin{table}[h!]
\centering
\begin{tabular}{|c||c|}
\hline
$a $ & $ \begin{array}{c}   \  \frac{ 4k^2 (N^2-1)-5  (k+N) \left(  \frac{(8-3\ell) }{5}N-2+\text{GCD}(k,N)  \right)}{48 (k+N)}  \\
+ \frac{ N }{8 (k+N)}   \sum_{j=1}^{N-1} \left\{\frac{j (k+N)}{N}  \right\} \left( 1 - \left\{\frac{j (k+N)}{N}  \right\} \right) \\ 
+ \frac{  4N^3(1 -\frac{1}{\ell}) \left( 2 k  + N (1 - \frac{1}{\ell})\right) }{48 (k+N)}     \end{array}$   \\ \hline
$c $ & $  \begin{array}{c} \  \frac{k^2 (N^2-1)-(k+N) \left(N (2-\ell)-2 +  \text{GCD}(k,N)\right)    }{12 (k+N)}  \\
 + \frac{ N^3(1-\frac{1}{\ell}) (2 k  + N (1 - \frac{1}{\ell}) ) }{ 12 (k+N) }  \end{array}$  \\ \hline
$N,k \rightarrow \infty$ & $a=c=\frac{N^2 \left( k + N(1-\frac{1}{\ell})\right)^2}{12 (k+N)}$  \\ \hline
\end{tabular}
\caption{The central charges of the $(A_{N-1}^{(N)}[k],Y_{{\ell}})$ theories.
  $\{ x \}=x - \lfloor x \rfloor$ denotes the fractional part. 
  \label{tab:tp1}}
\end{table}

When $\ell=1$ and $k/N$ is an integer, a Lagrangian description of the SCFT was obtained in \cite{Agarwal:2017roi,Benvenuti:2017bpg} (see also \cite{Maruyoshi:2016tqk,Maruyoshi:2016aim} for the case $N=2$). Using the dual Lagrangian, a set of $2^N-2$ Higgs branch operators  can be constructed, with dimension~\cite{Benvenuti:2017bpg}
	\begin{align}
	\Delta =  k- \frac{k}{N}\ .
	\end{align}
At large $N$, this exactly matches with the dimension of
the wrapped M2-brane operators $\cO_2$ in \eqref{wrapped_dim}, \eqref{wrapped_charges}.
We expect that the field-theory degeneracy factor $2^N-2$   could 
be understood on the gravity side by studying the possible ways in which
the M2-brane can end on the M5-brane source. Heuristically, 
we can picture the M2-brane worldvolume, which has a disk topology,
as the collapsed version of a multi-pronged
configuration that can have a boundary component on each of the $N$ M5-branes
independently. Since the  M2-brane must end on at least one of them,
 the  degeneracy is~$2^N-1$.
Notice the mismatch by one between the degeneracy in field theory and in gravity.
It would be interesting to sharpen this argument and 
to understand the origin of the additional decoupled mode, which we expect is  associated to the center-of-mass mode of the M5-brane source stack.

\section{Discussion}

We have proposed   gravity duals for the 4d $\cN =2$ SCFTs 
$(A_{N-1}^{(N)}[k],Y_{\ell})$
of AD type,
performing checks on the central charge, the $SU(\ell)$ flavor central charge,
and the dimensions of suitable Coulomb branch and Higgs branch operators.
Our $AdS_5$ solutions contain internal M5-brane sources.
They admit an isometry algebra $\mathfrak {su}(2)_R \oplus \mathfrak u(1)_r \oplus \mathfrak u(1)_\beta$. The $\mathfrak {su}(2)_R \oplus \mathfrak u(1)_r$ is dual to the SCFT R-symmetry,
while $u(1)_\beta$ does \emph{not} yield a continuous flavor symmetry thanks to a 
St\"uckelberg mechanism in which the $u(1)_\beta$ vector eats an axion originating   
from the expansion of the M-theory 3-form.
There could be still a discrete symmetry remnant of $\mathfrak u(1)_\beta$,
which we plan to study elsewhere.

We expect our 11d solutions to admit generalizations 
corresponding to a regular puncture labeled by an arbitrary Young diagram.
Constructing Lagrangian descriptions for these cases would yield
further insights into SCFTs of AD type and
allow for precision tests of the  holographic duality.

It would   be interesting to investigate whether the classification of irregular
punctures in field theory can be recovered by a systematic study of
exact solutions to the Toda equation of the class we discovered.

Our results set the stage for a broader   study of holographic duals of AD theories.  The supergravity constructions can be generalized to obtain $\cN=1$ systems.  More interestingly, our 
solutions can be used to study the holographic dual of the supersymmetry enhancing flows 
observed in the Lagrangian realizations of AD theories.

\begin{acknowledgments}
\emph{Acknowledgments.} We are grateful to 
Nikolay Bobev, Simone Giacomelli,
 Yifan Wang
 for interesting
conversations and correspondence. 
The work of IB   is supported in part by NSF grant PHY-1820784.
FB is supported by the European Union’s Horizon 2020 Framework: ERC Consolidator Grant 682608.
FB is supported by STFC Consolidated Grant ST/T000864/1.
RM is supported in part by ERC Grant 787320-QBH Structure
and by ERC Grant 772408-Stringlandscape.
The work of EN is supported by DOE grant DE-SC0020421.

\end{acknowledgments}

\bibliography{refs}

\begin{thebibliography}{41}%
\makeatletter
\providecommand \@ifxundefined [1]{%
 \@ifx{#1\undefined}
}%
\providecommand \@ifnum [1]{%
 \ifnum #1\expandafter \@firstoftwo
 \else \expandafter \@secondoftwo
 \fi
}%
\providecommand \@ifx [1]{%
 \ifx #1\expandafter \@firstoftwo
 \else \expandafter \@secondoftwo
 \fi
}%
\providecommand \natexlab [1]{#1}%
\providecommand \enquote  [1]{``#1''}%
\providecommand \bibnamefont  [1]{#1}%
\providecommand \bibfnamefont [1]{#1}%
\providecommand \citenamefont [1]{#1}%
\providecommand \href@noop [0]{\@secondoftwo}%
\providecommand \href [0]{\begingroup \@sanitize@url \@href}%
\providecommand \@href[1]{\@@startlink{#1}\@@href}%
\providecommand \@@href[1]{\endgroup#1\@@endlink}%
\providecommand \@sanitize@url [0]{\catcode `\\12\catcode `\$12\catcode
  `\&12\catcode `\#12\catcode `\^12\catcode `\_12\catcode `\%12\relax}%
\providecommand \@@startlink[1]{}%
\providecommand \@@endlink[0]{}%
\providecommand \url  [0]{\begingroup\@sanitize@url \@url }%
\providecommand \@url [1]{\endgroup\@href {#1}{\urlprefix }}%
\providecommand \urlprefix  [0]{URL }%
\providecommand \Eprint [0]{\href }%
\providecommand \doibase [0]{http://dx.doi.org/}%
\providecommand \selectlanguage [0]{\@gobble}%
\providecommand \bibinfo  [0]{\@secondoftwo}%
\providecommand \bibfield  [0]{\@secondoftwo}%
\providecommand \translation [1]{[#1]}%
\providecommand \BibitemOpen [0]{}%
\providecommand \bibitemStop [0]{}%
\providecommand \bibitemNoStop [0]{.\EOS\space}%
\providecommand \EOS [0]{\spacefactor3000\relax}%
\providecommand \BibitemShut  [1]{\csname bibitem#1\endcsname}%
\let\auto@bib@innerbib\@empty
\bibitem [{\citenamefont {Argyres}\ and\ \citenamefont
  {Douglas}(1995)}]{Argyres:1995jj}%
  \BibitemOpen
  \bibfield  {author} {\bibinfo {author} {\bibfnamefont {P.~C.}\ \bibnamefont
  {Argyres}}\ and\ \bibinfo {author} {\bibfnamefont {M.~R.}\ \bibnamefont
  {Douglas}},\ }\href {\doibase 10.1016/0550-3213(95)00281-V} {\bibfield
  {journal} {\bibinfo  {journal} {Nucl. Phys. B}\ }\textbf {\bibinfo {volume}
  {448}},\ \bibinfo {pages} {93} (\bibinfo {year} {1995})},\ \Eprint
  {http://arxiv.org/abs/hep-th/9505062} {arXiv:hep-th/9505062} \BibitemShut
  {NoStop}%
\bibitem [{\citenamefont {Agarwal}\ \emph {et~al.}(2017)\citenamefont
  {Agarwal}, \citenamefont {Sciarappa},\ and\ \citenamefont
  {Song}}]{Agarwal:2017roi}%
  \BibitemOpen
  \bibfield  {author} {\bibinfo {author} {\bibfnamefont {P.}~\bibnamefont
  {Agarwal}}, \bibinfo {author} {\bibfnamefont {A.}~\bibnamefont {Sciarappa}},
  \ and\ \bibinfo {author} {\bibfnamefont {J.}~\bibnamefont {Song}},\ }\href
  {\doibase 10.1007/JHEP10(2017)211} {\bibfield  {journal} {\bibinfo  {journal}
  {JHEP}\ }\textbf {\bibinfo {volume} {10}},\ \bibinfo {pages} {211} (\bibinfo
  {year} {2017})},\ \Eprint {http://arxiv.org/abs/1707.04751} {arXiv:1707.04751
  [hep-th]} \BibitemShut {NoStop}%
\bibitem [{\citenamefont {Benvenuti}\ and\ \citenamefont
  {Giacomelli}(2017)}]{Benvenuti:2017bpg}%
  \BibitemOpen
  \bibfield  {author} {\bibinfo {author} {\bibfnamefont {S.}~\bibnamefont
  {Benvenuti}}\ and\ \bibinfo {author} {\bibfnamefont {S.}~\bibnamefont
  {Giacomelli}},\ }\href {\doibase 10.1007/JHEP10(2017)106} {\bibfield
  {journal} {\bibinfo  {journal} {JHEP}\ }\textbf {\bibinfo {volume} {10}},\
  \bibinfo {pages} {106} (\bibinfo {year} {2017})},\ \Eprint
  {http://arxiv.org/abs/1707.05113} {arXiv:1707.05113 [hep-th]} \BibitemShut
  {NoStop}%
\bibitem [{\citenamefont {Brandhuber}\ and\ \citenamefont
  {Oz}(1999)}]{Brandhuber:1999np}%
  \BibitemOpen
  \bibfield  {author} {\bibinfo {author} {\bibfnamefont {A.}~\bibnamefont
  {Brandhuber}}\ and\ \bibinfo {author} {\bibfnamefont {Y.}~\bibnamefont
  {Oz}},\ }\href {\doibase 10.1016/S0370-2693(99)00763-7} {\bibfield  {journal}
  {\bibinfo  {journal} {Phys. Lett. B}\ }\textbf {\bibinfo {volume} {460}},\
  \bibinfo {pages} {307} (\bibinfo {year} {1999})},\ \Eprint
  {http://arxiv.org/abs/hep-th/9905148} {arXiv:hep-th/9905148} \BibitemShut
  {NoStop}%
\bibitem [{\citenamefont {Apruzzi}\ \emph {et~al.}(2014)\citenamefont
  {Apruzzi}, \citenamefont {Fazzi}, \citenamefont {Rosa},\ and\ \citenamefont
  {Tomasiello}}]{Apruzzi:2013yva}%
  \BibitemOpen
  \bibfield  {author} {\bibinfo {author} {\bibfnamefont {F.}~\bibnamefont
  {Apruzzi}}, \bibinfo {author} {\bibfnamefont {M.}~\bibnamefont {Fazzi}},
  \bibinfo {author} {\bibfnamefont {D.}~\bibnamefont {Rosa}}, \ and\ \bibinfo
  {author} {\bibfnamefont {A.}~\bibnamefont {Tomasiello}},\ }\href {\doibase
  10.1007/JHEP04(2014)064} {\bibfield  {journal} {\bibinfo  {journal} {JHEP}\
  }\textbf {\bibinfo {volume} {04}},\ \bibinfo {pages} {064} (\bibinfo {year}
  {2014})},\ \Eprint {http://arxiv.org/abs/1309.2949} {arXiv:1309.2949
  [hep-th]} \BibitemShut {NoStop}%
\bibitem [{\citenamefont {Gaiotto}\ and\ \citenamefont
  {Tomasiello}(2014)}]{Gaiotto:2014lca}%
  \BibitemOpen
  \bibfield  {author} {\bibinfo {author} {\bibfnamefont {D.}~\bibnamefont
  {Gaiotto}}\ and\ \bibinfo {author} {\bibfnamefont {A.}~\bibnamefont
  {Tomasiello}},\ }\href {\doibase 10.1007/JHEP12(2014)003} {\bibfield
  {journal} {\bibinfo  {journal} {JHEP}\ }\textbf {\bibinfo {volume} {12}},\
  \bibinfo {pages} {003} (\bibinfo {year} {2014})},\ \Eprint
  {http://arxiv.org/abs/1404.0711} {arXiv:1404.0711 [hep-th]} \BibitemShut
  {NoStop}%
\bibitem [{\citenamefont {Apruzzi}\ \emph {et~al.}(2015)\citenamefont
  {Apruzzi}, \citenamefont {Fazzi}, \citenamefont {Passias}, \citenamefont
  {Rota},\ and\ \citenamefont {Tomasiello}}]{Apruzzi:2015wna}%
  \BibitemOpen
  \bibfield  {author} {\bibinfo {author} {\bibfnamefont {F.}~\bibnamefont
  {Apruzzi}}, \bibinfo {author} {\bibfnamefont {M.}~\bibnamefont {Fazzi}},
  \bibinfo {author} {\bibfnamefont {A.}~\bibnamefont {Passias}}, \bibinfo
  {author} {\bibfnamefont {A.}~\bibnamefont {Rota}}, \ and\ \bibinfo {author}
  {\bibfnamefont {A.}~\bibnamefont {Tomasiello}},\ }\href {\doibase
  10.1103/PhysRevLett.115.061601} {\bibfield  {journal} {\bibinfo  {journal}
  {Phys. Rev. Lett.}\ }\textbf {\bibinfo {volume} {115}},\ \bibinfo {pages}
  {061601} (\bibinfo {year} {2015})},\ \Eprint
  {http://arxiv.org/abs/1502.06616} {arXiv:1502.06616 [hep-th]} \BibitemShut
  {NoStop}%
\bibitem [{\citenamefont {D'Hoker}\ \emph {et~al.}(2016)\citenamefont
  {D'Hoker}, \citenamefont {Gutperle}, \citenamefont {Karch},\ and\
  \citenamefont {Uhlemann}}]{DHoker:2016ujz}%
  \BibitemOpen
  \bibfield  {author} {\bibinfo {author} {\bibfnamefont {E.}~\bibnamefont
  {D'Hoker}}, \bibinfo {author} {\bibfnamefont {M.}~\bibnamefont {Gutperle}},
  \bibinfo {author} {\bibfnamefont {A.}~\bibnamefont {Karch}}, \ and\ \bibinfo
  {author} {\bibfnamefont {C.~F.}\ \bibnamefont {Uhlemann}},\ }\href {\doibase
  10.1007/JHEP08(2016)046} {\bibfield  {journal} {\bibinfo  {journal} {JHEP}\
  }\textbf {\bibinfo {volume} {08}},\ \bibinfo {pages} {046} (\bibinfo {year}
  {2016})},\ \Eprint {http://arxiv.org/abs/1606.01254} {arXiv:1606.01254
  [hep-th]} \BibitemShut {NoStop}%
\bibitem [{\citenamefont {D'Hoker}\ \emph {et~al.}(2017)\citenamefont
  {D'Hoker}, \citenamefont {Gutperle},\ and\ \citenamefont
  {Uhlemann}}]{DHoker:2017mds}%
  \BibitemOpen
  \bibfield  {author} {\bibinfo {author} {\bibfnamefont {E.}~\bibnamefont
  {D'Hoker}}, \bibinfo {author} {\bibfnamefont {M.}~\bibnamefont {Gutperle}}, \
  and\ \bibinfo {author} {\bibfnamefont {C.~F.}\ \bibnamefont {Uhlemann}},\
  }\href {\doibase 10.1007/JHEP05(2017)131} {\bibfield  {journal} {\bibinfo
  {journal} {JHEP}\ }\textbf {\bibinfo {volume} {05}},\ \bibinfo {pages} {131}
  (\bibinfo {year} {2017})},\ \Eprint {http://arxiv.org/abs/1703.08186}
  {arXiv:1703.08186 [hep-th]} \BibitemShut {NoStop}%
\bibitem [{\citenamefont {Bah}\ \emph {et~al.}(2017)\citenamefont {Bah},
  \citenamefont {Passias},\ and\ \citenamefont {Tomasiello}}]{Bah:2017wxp}%
  \BibitemOpen
  \bibfield  {author} {\bibinfo {author} {\bibfnamefont {I.}~\bibnamefont
  {Bah}}, \bibinfo {author} {\bibfnamefont {A.}~\bibnamefont {Passias}}, \ and\
  \bibinfo {author} {\bibfnamefont {A.}~\bibnamefont {Tomasiello}},\ }\href
  {\doibase 10.1007/JHEP11(2017)050} {\bibfield  {journal} {\bibinfo  {journal}
  {JHEP}\ }\textbf {\bibinfo {volume} {11}},\ \bibinfo {pages} {050} (\bibinfo
  {year} {2017})},\ \Eprint {http://arxiv.org/abs/1704.07389} {arXiv:1704.07389
  [hep-th]} \BibitemShut {NoStop}%
\bibitem [{\citenamefont {Bah}\ \emph {et~al.}(2019)\citenamefont {Bah},
  \citenamefont {Passias},\ and\ \citenamefont {Weck}}]{Bah:2018lyv}%
  \BibitemOpen
  \bibfield  {author} {\bibinfo {author} {\bibfnamefont {I.}~\bibnamefont
  {Bah}}, \bibinfo {author} {\bibfnamefont {A.}~\bibnamefont {Passias}}, \ and\
  \bibinfo {author} {\bibfnamefont {P.}~\bibnamefont {Weck}},\ }\href {\doibase
  10.1007/JHEP01(2019)058} {\bibfield  {journal} {\bibinfo  {journal} {JHEP}\
  }\textbf {\bibinfo {volume} {01}},\ \bibinfo {pages} {058} (\bibinfo {year}
  {2019})},\ \Eprint {http://arxiv.org/abs/1807.06031} {arXiv:1807.06031
  [hep-th]} \BibitemShut {NoStop}%
\bibitem [{\citenamefont {Bah}\ \emph {et~al.}()\citenamefont {Bah},
  \citenamefont {Bonetti}, \citenamefont {Minasian},\ and\ \citenamefont
  {Nardoni}}]{toappear}%
  \BibitemOpen
  \bibfield  {author} {\bibinfo {author} {\bibfnamefont {I.}~\bibnamefont
  {Bah}}, \bibinfo {author} {\bibfnamefont {F.}~\bibnamefont {Bonetti}},
  \bibinfo {author} {\bibfnamefont {R.}~\bibnamefont {Minasian}}, \ and\
  \bibinfo {author} {\bibfnamefont {E.}~\bibnamefont {Nardoni}},\ }\href@noop
  {} {\bibinfo  {journal} {to appear}\ }\BibitemShut {NoStop}%
\bibitem [{\citenamefont {Bah}\ \emph {et~al.}(2014)\citenamefont {Bah},
  \citenamefont {Gabella},\ and\ \citenamefont {Halmagyi}}]{Bah:2013wda}%
  \BibitemOpen
\bibfield  {journal} {  }\bibfield  {author} {\bibinfo {author} {\bibfnamefont
  {I.}~\bibnamefont {Bah}}, \bibinfo {author} {\bibfnamefont {M.}~\bibnamefont
  {Gabella}}, \ and\ \bibinfo {author} {\bibfnamefont {N.}~\bibnamefont
  {Halmagyi}},\ }\href {\doibase 10.1007/JHEP07(2014)131} {\bibfield  {journal}
  {\bibinfo  {journal} {JHEP}\ }\textbf {\bibinfo {volume} {07}},\ \bibinfo
  {pages} {131} (\bibinfo {year} {2014})},\ \Eprint
  {http://arxiv.org/abs/1312.6687} {arXiv:1312.6687 [hep-th]} \BibitemShut
  {NoStop}%
\bibitem [{\citenamefont {Ferrero}\ \emph {et~al.}(2021)\citenamefont
  {Ferrero}, \citenamefont {Gauntlett}, \citenamefont {P\'erez Ipi\~na},
  \citenamefont {Martelli},\ and\ \citenamefont {Sparks}}]{Ferrero:2020laf}%
  \BibitemOpen
  \bibfield  {author} {\bibinfo {author} {\bibfnamefont {P.}~\bibnamefont
  {Ferrero}}, \bibinfo {author} {\bibfnamefont {J.~P.}\ \bibnamefont
  {Gauntlett}}, \bibinfo {author} {\bibfnamefont {J.~M.}\ \bibnamefont {P\'erez
  Ipi\~na}}, \bibinfo {author} {\bibfnamefont {D.}~\bibnamefont {Martelli}}, \
  and\ \bibinfo {author} {\bibfnamefont {J.}~\bibnamefont {Sparks}},\ }\href
  {\doibase 10.1103/PhysRevLett.126.111601} {\bibfield  {journal} {\bibinfo
  {journal} {Phys. Rev. Lett.}\ }\textbf {\bibinfo {volume} {126}},\ \bibinfo
  {pages} {111601} (\bibinfo {year} {2021})},\ \Eprint
  {http://arxiv.org/abs/2011.10579} {arXiv:2011.10579 [hep-th]} \BibitemShut
  {NoStop}%
\bibitem [{\citenamefont {Ferrero}\ \emph {et~al.}(2020)\citenamefont
  {Ferrero}, \citenamefont {Gauntlett}, \citenamefont {Ipi\~na}, \citenamefont
  {Martelli},\ and\ \citenamefont {Sparks}}]{Ferrero:2020twa}%
  \BibitemOpen
  \bibfield  {author} {\bibinfo {author} {\bibfnamefont {P.}~\bibnamefont
  {Ferrero}}, \bibinfo {author} {\bibfnamefont {J.~P.}\ \bibnamefont
  {Gauntlett}}, \bibinfo {author} {\bibfnamefont {J.~M.~P.}\ \bibnamefont
  {Ipi\~na}}, \bibinfo {author} {\bibfnamefont {D.}~\bibnamefont {Martelli}}, \
  and\ \bibinfo {author} {\bibfnamefont {J.}~\bibnamefont {Sparks}},\
  }\href@noop {} {\  (\bibinfo {year} {2020})},\ \Eprint
  {http://arxiv.org/abs/2012.08530} {arXiv:2012.08530 [hep-th]} \BibitemShut
  {NoStop}%
\bibitem [{\citenamefont {Lin}\ \emph {et~al.}(2004)\citenamefont {Lin},
  \citenamefont {Lunin},\ and\ \citenamefont {Maldacena}}]{Lin:2004nb}%
  \BibitemOpen
  \bibfield  {author} {\bibinfo {author} {\bibfnamefont {H.}~\bibnamefont
  {Lin}}, \bibinfo {author} {\bibfnamefont {O.}~\bibnamefont {Lunin}}, \ and\
  \bibinfo {author} {\bibfnamefont {J.~M.}\ \bibnamefont {Maldacena}},\ }\href
  {\doibase 10.1088/1126-6708/2004/10/025} {\bibfield  {journal} {\bibinfo
  {journal} {JHEP}\ }\textbf {\bibinfo {volume} {10}},\ \bibinfo {pages} {025}
  (\bibinfo {year} {2004})},\ \Eprint {http://arxiv.org/abs/hep-th/0409174}
  {arXiv:hep-th/0409174} \BibitemShut {NoStop}%
\bibitem [{\citenamefont {Gaiotto}\ and\ \citenamefont
  {Maldacena}(2012)}]{Gaiotto:2009gz}%
  \BibitemOpen
  \bibfield  {author} {\bibinfo {author} {\bibfnamefont {D.}~\bibnamefont
  {Gaiotto}}\ and\ \bibinfo {author} {\bibfnamefont {J.}~\bibnamefont
  {Maldacena}},\ }\href {\doibase 10.1007/JHEP10(2012)189} {\bibfield
  {journal} {\bibinfo  {journal} {JHEP}\ }\textbf {\bibinfo {volume} {10}},\
  \bibinfo {pages} {189} (\bibinfo {year} {2012})},\ \Eprint
  {http://arxiv.org/abs/0904.4466} {arXiv:0904.4466 [hep-th]} \BibitemShut
  {NoStop}%
\bibitem [{\citenamefont {Maldacena}\ and\ \citenamefont
  {Nunez}(2001)}]{Maldacena:2000mw}%
  \BibitemOpen
  \bibfield  {author} {\bibinfo {author} {\bibfnamefont {J.~M.}\ \bibnamefont
  {Maldacena}}\ and\ \bibinfo {author} {\bibfnamefont {C.}~\bibnamefont
  {Nunez}},\ }\href {\doibase 10.1142/S0217751X01003937} {\bibfield  {journal}
  {\bibinfo  {journal} {Int. J. Mod. Phys. A}\ }\textbf {\bibinfo {volume}
  {16}},\ \bibinfo {pages} {822} (\bibinfo {year} {2001})},\ \Eprint
  {http://arxiv.org/abs/hep-th/0007018} {arXiv:hep-th/0007018} \BibitemShut
  {NoStop}%
\bibitem [{\citenamefont {Gauntlett}\ \emph {et~al.}(2007)\citenamefont
  {Gauntlett}, \citenamefont {O~Colgain},\ and\ \citenamefont
  {Varela}}]{Gauntlett:2006ai}%
  \BibitemOpen
  \bibfield  {author} {\bibinfo {author} {\bibfnamefont {J.~P.}\ \bibnamefont
  {Gauntlett}}, \bibinfo {author} {\bibfnamefont {E.}~\bibnamefont
  {O~Colgain}}, \ and\ \bibinfo {author} {\bibfnamefont {O.}~\bibnamefont
  {Varela}},\ }\href {\doibase 10.1088/1126-6708/2007/02/049} {\bibfield
  {journal} {\bibinfo  {journal} {JHEP}\ }\textbf {\bibinfo {volume} {02}},\
  \bibinfo {pages} {049} (\bibinfo {year} {2007})},\ \Eprint
  {http://arxiv.org/abs/hep-th/0611219} {arXiv:hep-th/0611219} \BibitemShut
  {NoStop}%
\bibitem [{\citenamefont {Bah}\ \emph {et~al.}(2020)\citenamefont {Bah},
  \citenamefont {Bonetti}, \citenamefont {Minasian},\ and\ \citenamefont
  {Nardoni}}]{Bah:2019rgq}%
  \BibitemOpen
  \bibfield  {author} {\bibinfo {author} {\bibfnamefont {I.}~\bibnamefont
  {Bah}}, \bibinfo {author} {\bibfnamefont {F.}~\bibnamefont {Bonetti}},
  \bibinfo {author} {\bibfnamefont {R.}~\bibnamefont {Minasian}}, \ and\
  \bibinfo {author} {\bibfnamefont {E.}~\bibnamefont {Nardoni}},\ }\href
  {\doibase 10.1007/JHEP01(2020)125} {\bibfield  {journal} {\bibinfo  {journal}
  {JHEP}\ }\textbf {\bibinfo {volume} {01}},\ \bibinfo {pages} {125} (\bibinfo
  {year} {2020})},\ \Eprint {http://arxiv.org/abs/1910.04166} {arXiv:1910.04166
  [hep-th]} \BibitemShut {NoStop}%
\bibitem [{\citenamefont {Bah}\ \emph {et~al.}(2021)\citenamefont {Bah},
  \citenamefont {Bonetti}, \citenamefont {Minasian},\ and\ \citenamefont
  {Nardoni}}]{Bah:2021hei}%
  \BibitemOpen
  \bibfield  {author} {\bibinfo {author} {\bibfnamefont {I.}~\bibnamefont
  {Bah}}, \bibinfo {author} {\bibfnamefont {F.}~\bibnamefont {Bonetti}},
  \bibinfo {author} {\bibfnamefont {R.}~\bibnamefont {Minasian}}, \ and\
  \bibinfo {author} {\bibfnamefont {E.}~\bibnamefont {Nardoni}},\ }\href@noop
  {} {\  (\bibinfo {year} {2021})},\ \Eprint {http://arxiv.org/abs/2106.01322}
  {arXiv:2106.01322 [hep-th]} \BibitemShut {NoStop}%
\bibitem [{\citenamefont {Wu}(1993)}]{WU1993381}%
  \BibitemOpen
  \bibfield  {author} {\bibinfo {author} {\bibfnamefont {S.}~\bibnamefont
  {Wu}},\ }\href {\doibase https://doi.org/10.1016/0393-0440(93)90005-Y}
  {\bibfield  {journal} {\bibinfo  {journal} {Journal of Geometry and Physics}\
  }\textbf {\bibinfo {volume} {10}},\ \bibinfo {pages} {381} (\bibinfo {year}
  {1993})}\BibitemShut {NoStop}%
\bibitem [{\citenamefont {Hinterbichler}\ \emph {et~al.}(2014)\citenamefont
  {Hinterbichler}, \citenamefont {Levin},\ and\ \citenamefont
  {Zukowski}}]{Hinterbichler:2013kwa}%
  \BibitemOpen
  \bibfield  {author} {\bibinfo {author} {\bibfnamefont {K.}~\bibnamefont
  {Hinterbichler}}, \bibinfo {author} {\bibfnamefont {J.}~\bibnamefont
  {Levin}}, \ and\ \bibinfo {author} {\bibfnamefont {C.}~\bibnamefont
  {Zukowski}},\ }\href {\doibase 10.1103/PhysRevD.89.086007} {\bibfield
  {journal} {\bibinfo  {journal} {Phys. Rev. D}\ }\textbf {\bibinfo {volume}
  {89}},\ \bibinfo {pages} {086007} (\bibinfo {year} {2014})},\ \Eprint
  {http://arxiv.org/abs/1310.6353} {arXiv:1310.6353 [hep-th]} \BibitemShut
  {NoStop}%
\bibitem [{\citenamefont {Xie}(2013)}]{Xie:2012hs}%
  \BibitemOpen
  \bibfield  {author} {\bibinfo {author} {\bibfnamefont {D.}~\bibnamefont
  {Xie}},\ }\href {\doibase 10.1007/JHEP01(2013)100} {\bibfield  {journal}
  {\bibinfo  {journal} {JHEP}\ }\textbf {\bibinfo {volume} {01}},\ \bibinfo
  {pages} {100} (\bibinfo {year} {2013})},\ \Eprint
  {http://arxiv.org/abs/1204.2270} {arXiv:1204.2270 [hep-th]} \BibitemShut
  {NoStop}%
\bibitem [{\citenamefont {Wang}\ and\ \citenamefont
  {Xie}(2016)}]{Wang:2015mra}%
  \BibitemOpen
  \bibfield  {author} {\bibinfo {author} {\bibfnamefont {Y.}~\bibnamefont
  {Wang}}\ and\ \bibinfo {author} {\bibfnamefont {D.}~\bibnamefont {Xie}},\
  }\href {\doibase 10.1103/PhysRevD.94.065012} {\bibfield  {journal} {\bibinfo
  {journal} {Phys. Rev. D}\ }\textbf {\bibinfo {volume} {94}},\ \bibinfo
  {pages} {065012} (\bibinfo {year} {2016})},\ \Eprint
  {http://arxiv.org/abs/1509.00847} {arXiv:1509.00847 [hep-th]} \BibitemShut
  {NoStop}%
\bibitem [{\citenamefont {Xie}\ and\ \citenamefont {Zhao}(2013)}]{Xie:2013jc}%
  \BibitemOpen
  \bibfield  {author} {\bibinfo {author} {\bibfnamefont {D.}~\bibnamefont
  {Xie}}\ and\ \bibinfo {author} {\bibfnamefont {P.}~\bibnamefont {Zhao}},\
  }\href {\doibase 10.1007/JHEP03(2013)006} {\bibfield  {journal} {\bibinfo
  {journal} {JHEP}\ }\textbf {\bibinfo {volume} {03}},\ \bibinfo {pages} {006}
  (\bibinfo {year} {2013})},\ \Eprint {http://arxiv.org/abs/1301.0210}
  {arXiv:1301.0210 [hep-th]} \BibitemShut {NoStop}%
\bibitem [{\citenamefont {Cecotti}\ \emph {et~al.}(2010)\citenamefont
  {Cecotti}, \citenamefont {Neitzke},\ and\ \citenamefont
  {Vafa}}]{Cecotti:2010fi}%
  \BibitemOpen
  \bibfield  {author} {\bibinfo {author} {\bibfnamefont {S.}~\bibnamefont
  {Cecotti}}, \bibinfo {author} {\bibfnamefont {A.}~\bibnamefont {Neitzke}}, \
  and\ \bibinfo {author} {\bibfnamefont {C.}~\bibnamefont {Vafa}},\ }\href@noop
  {} {\  (\bibinfo {year} {2010})},\ \Eprint {http://arxiv.org/abs/1006.3435}
  {arXiv:1006.3435 [hep-th]} \BibitemShut {NoStop}%
\bibitem [{\citenamefont {Argyres}\ \emph {et~al.}(1996)\citenamefont
  {Argyres}, \citenamefont {Plesser}, \citenamefont {Seiberg},\ and\
  \citenamefont {Witten}}]{Argyres:1995xn}%
  \BibitemOpen
  \bibfield  {author} {\bibinfo {author} {\bibfnamefont {P.~C.}\ \bibnamefont
  {Argyres}}, \bibinfo {author} {\bibfnamefont {M.}~\bibnamefont {Plesser}},
  \bibinfo {author} {\bibfnamefont {N.}~\bibnamefont {Seiberg}}, \ and\
  \bibinfo {author} {\bibfnamefont {E.}~\bibnamefont {Witten}},\ }\href
  {\doibase 10.1016/0550-3213(95)00671-0} {\bibfield  {journal} {\bibinfo
  {journal} {Nucl. Phys. B}\ }\textbf {\bibinfo {volume} {461}},\ \bibinfo
  {pages} {71} (\bibinfo {year} {1996})},\ \Eprint
  {http://arxiv.org/abs/hep-th/9511154} {arXiv:hep-th/9511154} \BibitemShut
  {NoStop}%
\bibitem [{\citenamefont {Eguchi}\ \emph {et~al.}(1996)\citenamefont {Eguchi},
  \citenamefont {Hori}, \citenamefont {Ito},\ and\ \citenamefont
  {Yang}}]{Eguchi:1996vu}%
  \BibitemOpen
  \bibfield  {author} {\bibinfo {author} {\bibfnamefont {T.}~\bibnamefont
  {Eguchi}}, \bibinfo {author} {\bibfnamefont {K.}~\bibnamefont {Hori}},
  \bibinfo {author} {\bibfnamefont {K.}~\bibnamefont {Ito}}, \ and\ \bibinfo
  {author} {\bibfnamefont {S.-K.}\ \bibnamefont {Yang}},\ }\href {\doibase
  10.1016/0550-3213(96)00188-5} {\bibfield  {journal} {\bibinfo  {journal}
  {Nucl. Phys. B}\ }\textbf {\bibinfo {volume} {471}},\ \bibinfo {pages} {430}
  (\bibinfo {year} {1996})},\ \Eprint {http://arxiv.org/abs/hep-th/9603002}
  {arXiv:hep-th/9603002} \BibitemShut {NoStop}%
\bibitem [{Note1()}]{Note1}%
  \BibitemOpen
  \bibinfo {note} {Box Young diagrams from orbifold singularities were studied
  in \cite {Bobev:2019ore}.}\BibitemShut {Stop}%
\bibitem [{\citenamefont {Cecotti}\ and\ \citenamefont
  {Del~Zotto}(2013)}]{Cecotti:2012jx}%
  \BibitemOpen
  \bibfield  {author} {\bibinfo {author} {\bibfnamefont {S.}~\bibnamefont
  {Cecotti}}\ and\ \bibinfo {author} {\bibfnamefont {M.}~\bibnamefont
  {Del~Zotto}},\ }\href {\doibase 10.1007/JHEP01(2013)191} {\bibfield
  {journal} {\bibinfo  {journal} {JHEP}\ }\textbf {\bibinfo {volume} {01}},\
  \bibinfo {pages} {191} (\bibinfo {year} {2013})},\ \Eprint
  {http://arxiv.org/abs/1210.2886} {arXiv:1210.2886 [hep-th]} \BibitemShut
  {NoStop}%
\bibitem [{\citenamefont {Cecotti}\ \emph {et~al.}(2013)\citenamefont
  {Cecotti}, \citenamefont {Del~Zotto},\ and\ \citenamefont
  {Giacomelli}}]{Cecotti:2013lda}%
  \BibitemOpen
  \bibfield  {author} {\bibinfo {author} {\bibfnamefont {S.}~\bibnamefont
  {Cecotti}}, \bibinfo {author} {\bibfnamefont {M.}~\bibnamefont {Del~Zotto}},
  \ and\ \bibinfo {author} {\bibfnamefont {S.}~\bibnamefont {Giacomelli}},\
  }\href {\doibase 10.1007/JHEP04(2013)153} {\bibfield  {journal} {\bibinfo
  {journal} {JHEP}\ }\textbf {\bibinfo {volume} {04}},\ \bibinfo {pages} {153}
  (\bibinfo {year} {2013})},\ \Eprint {http://arxiv.org/abs/1303.3149}
  {arXiv:1303.3149 [hep-th]} \BibitemShut {NoStop}%
\bibitem [{\citenamefont {Giacomelli}(2018)}]{Giacomelli:2017ckh}%
  \BibitemOpen
  \bibfield  {author} {\bibinfo {author} {\bibfnamefont {S.}~\bibnamefont
  {Giacomelli}},\ }\href {\doibase 10.1007/JHEP06(2018)156} {\bibfield
  {journal} {\bibinfo  {journal} {JHEP}\ }\textbf {\bibinfo {volume} {06}},\
  \bibinfo {pages} {156} (\bibinfo {year} {2018})},\ \Eprint
  {http://arxiv.org/abs/1710.06469} {arXiv:1710.06469 [hep-th]} \BibitemShut
  {NoStop}%
\bibitem [{\citenamefont {Xie}\ and\ \citenamefont {Yau}(2015)}]{Xie:2015rpa}%
  \BibitemOpen
  \bibfield  {author} {\bibinfo {author} {\bibfnamefont {D.}~\bibnamefont
  {Xie}}\ and\ \bibinfo {author} {\bibfnamefont {S.-T.}\ \bibnamefont {Yau}},\
  }\href@noop {} {\  (\bibinfo {year} {2015})},\ \Eprint
  {http://arxiv.org/abs/1510.01324} {arXiv:1510.01324 [hep-th]} \BibitemShut
  {NoStop}%
\bibitem [{\citenamefont {Xie}\ \emph {et~al.}(2021)\citenamefont {Xie},
  \citenamefont {Yan},\ and\ \citenamefont {Yau}}]{Xie:2016evu}%
  \BibitemOpen
  \bibfield  {author} {\bibinfo {author} {\bibfnamefont {D.}~\bibnamefont
  {Xie}}, \bibinfo {author} {\bibfnamefont {W.}~\bibnamefont {Yan}}, \ and\
  \bibinfo {author} {\bibfnamefont {S.-T.}\ \bibnamefont {Yau}},\ }\href
  {\doibase 10.1103/PhysRevD.103.065003} {\bibfield  {journal} {\bibinfo
  {journal} {Phys. Rev. D}\ }\textbf {\bibinfo {volume} {103}},\ \bibinfo
  {pages} {065003} (\bibinfo {year} {2021})},\ \Eprint
  {http://arxiv.org/abs/1604.02155} {arXiv:1604.02155 [hep-th]} \BibitemShut
  {NoStop}%
\bibitem [{\citenamefont {Shapere}\ and\ \citenamefont
  {Tachikawa}(2008)}]{Shapere:2008zf}%
  \BibitemOpen
  \bibfield  {author} {\bibinfo {author} {\bibfnamefont {A.~D.}\ \bibnamefont
  {Shapere}}\ and\ \bibinfo {author} {\bibfnamefont {Y.}~\bibnamefont
  {Tachikawa}},\ }\href {\doibase 10.1088/1126-6708/2008/09/109} {\bibfield
  {journal} {\bibinfo  {journal} {JHEP}\ }\textbf {\bibinfo {volume} {09}},\
  \bibinfo {pages} {109} (\bibinfo {year} {2008})},\ \Eprint
  {http://arxiv.org/abs/0804.1957} {arXiv:0804.1957 [hep-th]} \BibitemShut
  {NoStop}%
\bibitem [{\citenamefont {Giacomelli}\ \emph {et~al.}(2021)\citenamefont
  {Giacomelli}, \citenamefont {Mekareeya},\ and\ \citenamefont
  {Sacchi}}]{Giacomelli:2020ryy}%
  \BibitemOpen
  \bibfield  {author} {\bibinfo {author} {\bibfnamefont {S.}~\bibnamefont
  {Giacomelli}}, \bibinfo {author} {\bibfnamefont {N.}~\bibnamefont
  {Mekareeya}}, \ and\ \bibinfo {author} {\bibfnamefont {M.}~\bibnamefont
  {Sacchi}},\ }\href {\doibase 10.1007/JHEP03(2021)242} {\bibfield  {journal}
  {\bibinfo  {journal} {JHEP}\ }\textbf {\bibinfo {volume} {03}},\ \bibinfo
  {pages} {242} (\bibinfo {year} {2021})},\ \Eprint
  {http://arxiv.org/abs/2012.12852} {arXiv:2012.12852 [hep-th]} \BibitemShut
  {NoStop}%
\bibitem [{Note2()}]{Note2}%
  \BibitemOpen
  \bibinfo {note} {A Coulomb branch operator has $U(1)_r$ charge $r = 2\Delta $
  and is uncharged under $SU(2)_R$.}\BibitemShut {Stop}%
\bibitem [{\citenamefont {Maruyoshi}\ and\ \citenamefont
  {Song}(2017{\natexlab{a}})}]{Maruyoshi:2016tqk}%
  \BibitemOpen
  \bibfield  {author} {\bibinfo {author} {\bibfnamefont {K.}~\bibnamefont
  {Maruyoshi}}\ and\ \bibinfo {author} {\bibfnamefont {J.}~\bibnamefont
  {Song}},\ }\href {\doibase 10.1103/PhysRevLett.118.151602} {\bibfield
  {journal} {\bibinfo  {journal} {Phys. Rev. Lett.}\ }\textbf {\bibinfo
  {volume} {118}},\ \bibinfo {pages} {151602} (\bibinfo {year}
  {2017}{\natexlab{a}})},\ \Eprint {http://arxiv.org/abs/1606.05632}
  {arXiv:1606.05632 [hep-th]} \BibitemShut {NoStop}%
\bibitem [{\citenamefont {Maruyoshi}\ and\ \citenamefont
  {Song}(2017{\natexlab{b}})}]{Maruyoshi:2016aim}%
  \BibitemOpen
  \bibfield  {author} {\bibinfo {author} {\bibfnamefont {K.}~\bibnamefont
  {Maruyoshi}}\ and\ \bibinfo {author} {\bibfnamefont {J.}~\bibnamefont
  {Song}},\ }\href {\doibase 10.1007/JHEP02(2017)075} {\bibfield  {journal}
  {\bibinfo  {journal} {JHEP}\ }\textbf {\bibinfo {volume} {02}},\ \bibinfo
  {pages} {075} (\bibinfo {year} {2017}{\natexlab{b}})},\ \Eprint
  {http://arxiv.org/abs/1607.04281} {arXiv:1607.04281 [hep-th]} \BibitemShut
  {NoStop}%
\bibitem [{\citenamefont {Bobev}\ \emph {et~al.}(2020)\citenamefont {Bobev},
  \citenamefont {Bomans},\ and\ \citenamefont {Gautason}}]{Bobev:2019ore}%
  \BibitemOpen
  \bibfield  {author} {\bibinfo {author} {\bibfnamefont {N.}~\bibnamefont
  {Bobev}}, \bibinfo {author} {\bibfnamefont {P.}~\bibnamefont {Bomans}}, \
  and\ \bibinfo {author} {\bibfnamefont {F.~F.}\ \bibnamefont {Gautason}},\
  }\href {\doibase 10.1007/JHEP06(2020)011} {\bibfield  {journal} {\bibinfo
  {journal} {JHEP}\ }\textbf {\bibinfo {volume} {06}},\ \bibinfo {pages} {011}
  (\bibinfo {year} {2020})},\ \Eprint {http://arxiv.org/abs/1912.04779}
  {arXiv:1912.04779 [hep-th]} \BibitemShut {NoStop}%
\end{thebibliography}%

\end{document}